\documentclass[conference,letterpaper,10pt]{IEEEtran}

\usepackage[english]{babel}
\usepackage[T1]{fontenc}
\usepackage[noadjust]{cite}
\usepackage{url,color} 
\usepackage{graphics,amsfonts}
\usepackage{epstopdf}
\usepackage[pdftex]{graphicx}
\usepackage[cmex10]{amsmath}
\usepackage[utf8]{inputenc}

\usepackage{times}

\usepackage{xcolor}

\makeatletter
\DeclareRobustCommand{\change}{%
  \@bsphack
  \leavevmode
  \color{red}%
  \@esphack
}
\DeclareRobustCommand{\stopchange}{%
  \@bsphack
  \normalcolor
  \@esphack
}
\makeatother

\begin{document}
\title{Game Theoretic Analysis of Road User Safety Scenarios Involving Autonomous Vehicles}

\author{\IEEEauthorblockN{Umberto Michieli,
Leonardo Badia}
\IEEEauthorblockA{Department of Information Engineering, University of Padova -- Via Gradenigo, 6/b, 35131 Padova, Italy\\Email: {\tt\{michieli, badia\}@dei.unipd.it}
}}

\maketitle

\begin{abstract}

Interactions between pedestrians, bikers, and human-driven vehicles have been a major concern in traffic safety over the years. The upcoming age of autonomous vehicles will further raise major problems on whether self-driving cars can accurately avoid accidents; on the other hand, usability issues arise on whether human-driven cars and pedestrians can dominate the road at the expense of the autonomous vehicles which will be programmed to avoid accidents. This paper proposes some game theoretical models applied to related traffic scenarios, where the strategic interaction between a pedestrian and an autonomous vehicle is analyzed. The games have been simulated in order to demonstrate the theoretical analysis and the predicted behaviors. These investigations can shed new lights on how novel urban traffic regulations could be required to allow for a better interaction of vehicles and a general improved management of traffic and communication vehicular networks.
\end{abstract}

\section{Introduction}\label{sec:intro}

\IEEEPARstart{S}{mart} connected machines are expected to be one of the most sensational innovations developed over the next ten years. Present-day technologies can benefit from the computational improvement of processors and the increase in the amount of data collected by smart sensors, as well as the empowerment of machine learning techniques such as deep neural networks, as was already pointed out in 2010 \cite{campbell}. The trend in this field will be likely led by autonomous vehicles (AVs), which might set up the most exciting revolution in everyday's customs and lifestyle of contemporary society. Connected and self-driving cars are currently set by the 2016 Gartner's Hype Cycle on the top of the \textit{Peak of Inflated Expectations} \cite{gartner}.

This can benefit the quality of life of citizens in future societies by several means. For example, AVs are expected to significantly reduce road accidents, since they can be less error-prone than the human drivers and be programmed to avoid unnecessary risks. We recall that in the European Union alone there are more than $25$ thousand deaths in road accidents every year \cite{europe}. Still, the human factor will be present in the interaction with pedestrian; thus, besides being risk averse per se, connected vehicles should also implement additional mechanisms, for example, they might also notify their presence to distracted pedestrians, as suggested in \cite{hussein}. 

An open challenge still involves the description of interactions in this future scenarios where human drivers and AVs, as well as pedestrians and cyclists, will coexist. Many studies have been proposed in the literature to shed light on these aspects, and game theoretical and/or statistical approaches have been advanced. Surveys and predictions of the future transport systems can be found, for example, in \cite{milakis} and \cite{millard}. The latter also gained mass attention thanks to popular divulgative versions such as \cite{hurley} and \cite{will}. Specifically, the idea of \cite{millard} is to build up a two-players game between a pedestrian crossing a street and a (possibly self-driven) vehicle reaching the pedestrian crossing. The actions of the players in face of a possible collision are either to keep moving in the intended direction or to yield. This analysis led to some considerations about possible behaviors of the pedestrian player, which can be summarized as concluding that pedestrians will be unsure on whether to cross the street or yield if they assume that the incoming vehicle is human-driven, since they may fear that the human driver will not stop, while they will more boldly cross the street in front of an AV, relying on it to stop. Thus, the author suggests that beyond increasing safety for pedestrians, a new scenario for coexistence of road users will open up where pedestrians will have the supremacy over the AVs and it may be necessary to regulate the traffic response.

Inspired by this open issue, in this paper we present the following contributions, where our goal is to review multiple road scenarios, all involving pedestrians and vehicles but including different points of view than the aforementioned one.
In this context, we will use a Bayesian game theoretical model to derive conclusions that will also be verified by computer simulation. More specifically, we will discuss first of all the impact of the increase in the share of vehicles that are connected to the Internet and autonomously driven in a classic scenario involving a pedestrian/cyclist crossing a street and an incoming vehicle. Thus, beyond generally concluding that the pedestrian player will eventually be safer and at the same time will get control of the road over the AVs, we will also be able to provide a quantitative evaluation of this phenomenon.
Furthermore, we will analyze the behavior of pedestrians in a more detailed situation where they can decide whether or not to cross, involving different parameters settings.

The remainder of this paper is organized as follows. Section \ref{sec:related} reviews the applications of game theory to road user scenarios already existing in the literature. Sections \ref{sec:res1} and \ref{sec:res2} presents two different scenarios, namely a Bayesian model for a Cyclist/Autonomous Vehicle interaction and a Pedestrian crossing modeled as a Bayesian entry game, respectiely, and discuss the obtained results. Section \ref{sec:conclusion} concludes the paper and outlines the main future guidelines and possible expansions for future work.

\section{Related Work}\label{sec:related}

Among the many applications of game theory,
in the last few years, an increasing interest of the scientific community was produced
towards the implementation of game theoretical techniques in traffic scenarios. 
Indeed, traffic interactions are a quite ideal field of application for game theory,
as they usually involve: conflicts with different perspectives of the users (and, as a result,
different utilities); a generally limited but clearly distinguishable set of options for
the involved players; occurrence of repetitions in many instances of the game,
thus allowing for large numbers and statistical generality.
Therefore, game theory can offer an accurate strategic analysis that can
be used to clarify whether certain road users have common or conflicting interests; it also shows explicitly the choices facing road users and is able to make specific predictions about road user behaviors, which are
therefore empirically testable.

In \cite{bjornskau}, Bj{\o}rnskau performed some interesting investigations about a \textit{Zebra Crossing Game} for the cyclists in Norway; he modeled the crossing game of a cyclist versus a driver where the cyclist does the first move and can yield, cycle, or dismount from the bike and walk, while the driver can either drive or yield. By applying backward induction, it can be shown that the only stable and perfect equilibrium turns out to be \textit{cycle} for the cyclist and \textit{yield} for the driver; of course the cyclist has the clear advantage of being the first mover; in other words, he is the leader of a Stackelberg game \cite{osborne}.

Interesting reviews of game theoretical analysis applied on road user behaviors can be found both in \cite{bjornskau1}, which is one of the first works trying to explain the complex dynamics involved in traffic accidents, and in \cite{elvik}, where a set of many road scenarios involving vehicles is proposed and solved.

A more sophisticated study about the micro-dynamics viewpoint of the interactions between vehicles and pedestrians is analyzed in \cite{pchen}, which models and simulates the relative speed between the pedestrian and the approaching car at uncontrolled mid-block crosswalks using a game theoretical approach.
Game theory was also exploited in \cite{rane} to perform conflict resolution of AVs: they provide a simulation scenario involving two crossing cars, which is further analyzed using Cellular Automata.
A pedestrian's decision model for determining the behavior of two pedestrians attempting to avoid collision when approaching a blind corner from both sides was studied in \cite{okamoto} exploiting a combination of game theory and the theory of evidence in order to account for uncertainty.

Others non-game-theory-based works that provide some simulation results are mainly based on statistical models. Those are not always able to capture the specific situations and the causal relationships, as stated in \cite{hauer}. For example in \cite{bchen} the interaction between the AVs and the pedestrians is stochastically modeled through a multivariate Gaussian function using real traffic data. Moreover, statistical tools are applied in \cite{schneemann}, where a scenario of automated vehicles and pedestrians at crosswalks is analyzed in order to show a significant influence of the vehicle speed on the pedestrian's decision process.

Our study brings the novel contribution of using more advanced game theory techniques than those used in the existing literature. Usually, game theory is employed as a ready-to-use tool, which gives it a marginal role to the game theoretic model in itself.
Instead, we will consider a Bayesian approach with a strong emphasis on the characterization of different types of the involved players: e.g., probability of a vehicle being human-driven or autonomous, as seen from the perspective of a pedestrian/cyclists that has unclear information and must resort to a prior probability.
Moreover, we will infer a behavioral characterization from numerical evidence, as it is reasonable to expect that, as the share of AVs increase, it will be more likely to encounter them and the human cyclist or pedestrian can adapt their prior to reflect that. This would also allow us to drive specific quantitative conclusions that can give useful numerical insights.



\section{Cyclist versus Vehicle using Bayesian Game}
\label{sec:res1}

The first scenario presented in this study regards the zebra crossing of a cyclist with an upcoming driver, which can either be autonomous or human-driven (and the cyclist tries to detect it). As in the scenario proposed by \cite{bjornskau}, the cyclist has three options in the strategy set: to yield, to dismount from the bike and walk, or to cycle. Let these actions be denoted as Y, W, C, respectively. On the other side, the vehicle can either go or stop and let the cyclist cross (G or S, respectively).
The street regulations generally state that zebra crossing are for pedestrian only; cyclists cannot cross the road there, but instead they should dismount from the bike and walk in order to have the right to cross. At the same time, the driver is supposed by the same regulations to stop and let the cyclist cross only if they has dismounted from the bike.

\begin{figure}[b!h]
\hspace{-0.2cm}\includegraphics[width=0.95\linewidth]{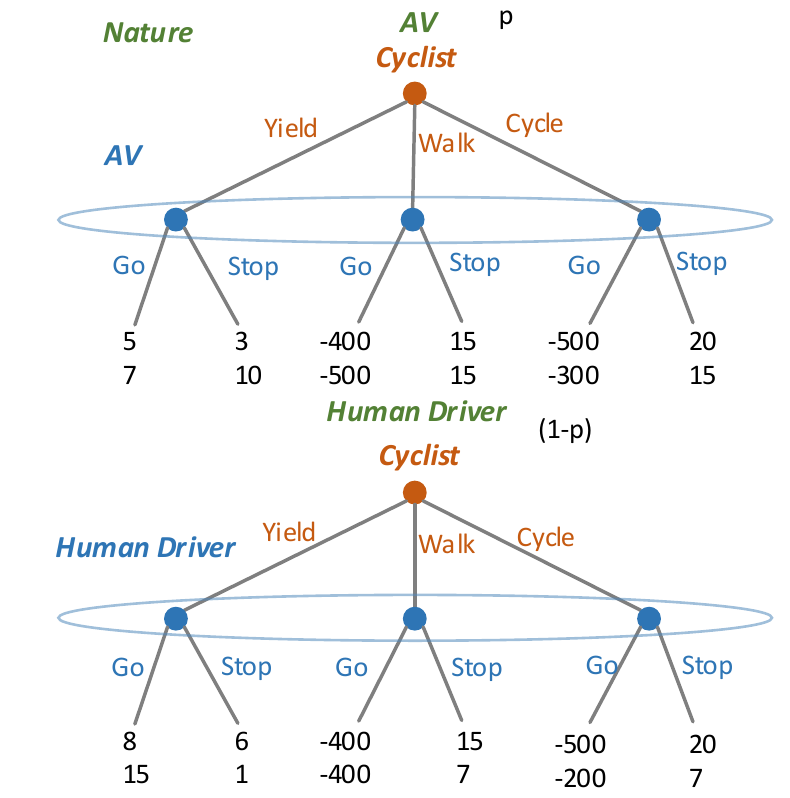}
\caption{Extensive form of the game in Section \ref{sec:res1} considering medium speed, payoffs of the cyclist are given first.}
\label{fig:1a}
\end{figure}

The situation described above can be modeled through a simultaneous Bayesian game with two types as depicted in extensive form in Fig.\ \ref{fig:1a} where the payoffs are set for a vehicle oncoming at a medium speed.
As customary in Bayesian game analysis \cite{osborne} a virtual player, ``nature,'' initially draws at random according to a prior distribution the character of the vehicle (i.e., whether it is autonomous or human-driven, with probabilities $p$ and $1-p$, respectively). The game is modeled as simultaneous because is considered the extreme case of last chance for both players to act. The novel element of our tractation is the Bayesian aspect that allows to take into account the uncertainty on the type of the vehicle.

Even though road users have to take important decisions in fractions of second, we can assume that common knowledge and full rationality still apply: the players are fully informed about the rules of the game, the possible actions, and the evaluation of each other player to all the outcomes.

The aim of this game is to compare the cyclist's behavior when facing an autonomous or a human-driven vehicle and to verify whether the accident rate will be reduced by increasing the number of autonomous vehicles; in principle, autonomous car can be programmed to be prompter to prevent accidents; but on the other hand, human road users can act more boldly in the presence of AVs.
The standard solutions in game theory are Nash Equilibria (NEs), which can be in either \textit{pure} or \textit{mixed} strategies, the latter case being seen as providing a probability distribution over the set of pure strategies of the players.
Since this game is also concerned about the accident rate, the \textit{mixed} NEs should be inspected: thus, the payoffs need to be assigned following a cardinality relation. Considering only the \textit{pure} ones, in fact, the players will always choose the same action, which is not interesting for the desired purpose.

It is assumed that both players prefer not to collide (as their primary goal) and secondly whether they can continue to move, which, if available, is a dominant outcome over having to stop and wait.\\
Summarizing, it is desirable that the payoffs reflect a heavy penalization in case of collisions (high absolute value for negative payoffs) and at the same time they do not overestimate the daily-action of crossing roads (low absolute value for positive payoffs).

Beyond these assumptions, the evaluation of the payoffs are challenging. The worst outcome for the cyclist (arbitrarily set to $-500$) happens when he crosses the road and is hit by the vehicle, no matter whether autonomous or human-driven, it is worse than to be hit when walking (hence set, for instance, to $-400$) because in this last case he is right according to the traffic regulations. Then, the next worst outcome for the cyclist is when he and the AV stop (slightly positive because he acted following the road rules, e.g. set to $3$), here they need to play again in order to solve the game, this is considered worse than to yield to an AV which does not stop ($5$). Same considerations apply in case of a human-driven vehicle, which leads to higher payoffs (respectively $6$ and $8$) because the cyclist knows that the AV is programmed to be \textit{risk averse} as much as possible while the driver can be absent-minded. The best outcomes for the cyclist are respectively to cycle over the zebra crossing with a yielding vehicle ($20$) and to dismount from the bike and walk over the zebra crossing with a yielding vehicle ($15$).

The worst outcome for the driver is an AV hitting a pedestrian (quantified as $-500$) because it is a machine-type evaluation error, which is considered to be worse and less socially accepted than a human-type error (quantified as $-400$). Same considerations apply in case of a crossing cyclist (with payoffs for AV and human-driver respectively $-300$ and $-200$): the only difference is that the cyclist is wrong, so the payoffs of the vehicle are higher than the previous case. The next worst outcome for the driver is when they both yield and it is a human driver ($1$), it is worse than the situation in which both the autonomous vehicle and the cyclist yield ($10$) because the AV is programmed to avoid accidents. This last situation is slightly preferable to a yielding human-driver when a cyclist cycles over the zebra crossing ($7$); this outcome is very similar to a yielding human-driver when the pedestrian walks and to an AV keeping driving when the cyclist yields, which is not as good as a yielding AV because it is \textit{risk averse}. The best outcome for the vehicle happens when an AV stops in front of a crossing cyclist because it is the right action for the AV and when an human-driven vehicle keeps going as the cyclist yields and stops.

In this game, there are two pure NEs: (CY, SG) and (CC, SS). Those can be found simplifying the game via \textit{Iterated Elimination of Strictly Dominated Strategies} (IESDS) \cite{osborne} and then seeking for the NEs. IESDS can be applied since all players are rational, i.e. they do not play strictly dominated strategies, and rationality is a common knowledge: hence the game is smaller and the procedure can be iterated.

The meaning of the two NEs found is: when the vehicle is autonomous then the cyclist knows that can cycle therefore the AV stops (\textit{stop} for the AV is a strictly dominant strategy), when the driver is human then both actions are available to the driver and the cyclist yields or cycles. In both cases no player has an incentive to deviate by changing his choice given the choices of the others.

The unique mixed NE is to always play (C,S) in case of autonomous vehicle and to mix between yield and cycle (with probabilities $207/221$ and $14/221$) and between go and stop (with probabilities $7/261$ and $254/261$) in case of human driver. These results have been obtained by means of computer simulations. Quite logically, the absolute numerical values depend by the payoffs chosen and will be modified if other values are set. Still, from a general perspective we can infer that, whatever the specific values chosen, for most of the times the players will choose to yield because the risk is too high; however, sometimes they can decide to go. 

In order to see that the rising number of AVs will decrease the number of collisions, a simulation of the mixed strategies was performed varying the probability of encountering an AV. In addition to that, the simulation was also repeated using other two different speed values of the oncoming vehicle. The new payoffs are computed on the basis of the ones considered in Fig.\ \ref{fig:1a}; in particular, for example, the payoffs of the cyclist would get lower in case of a vehicle driving at a higher speed because it would be a more serious accident and vice versa. Furthermore the robustness of the model has been evaluated by means of noisy payoffs thus confirming the general trend obtained. 

\begin{figure}
\centering
\includegraphics[width=\linewidth]{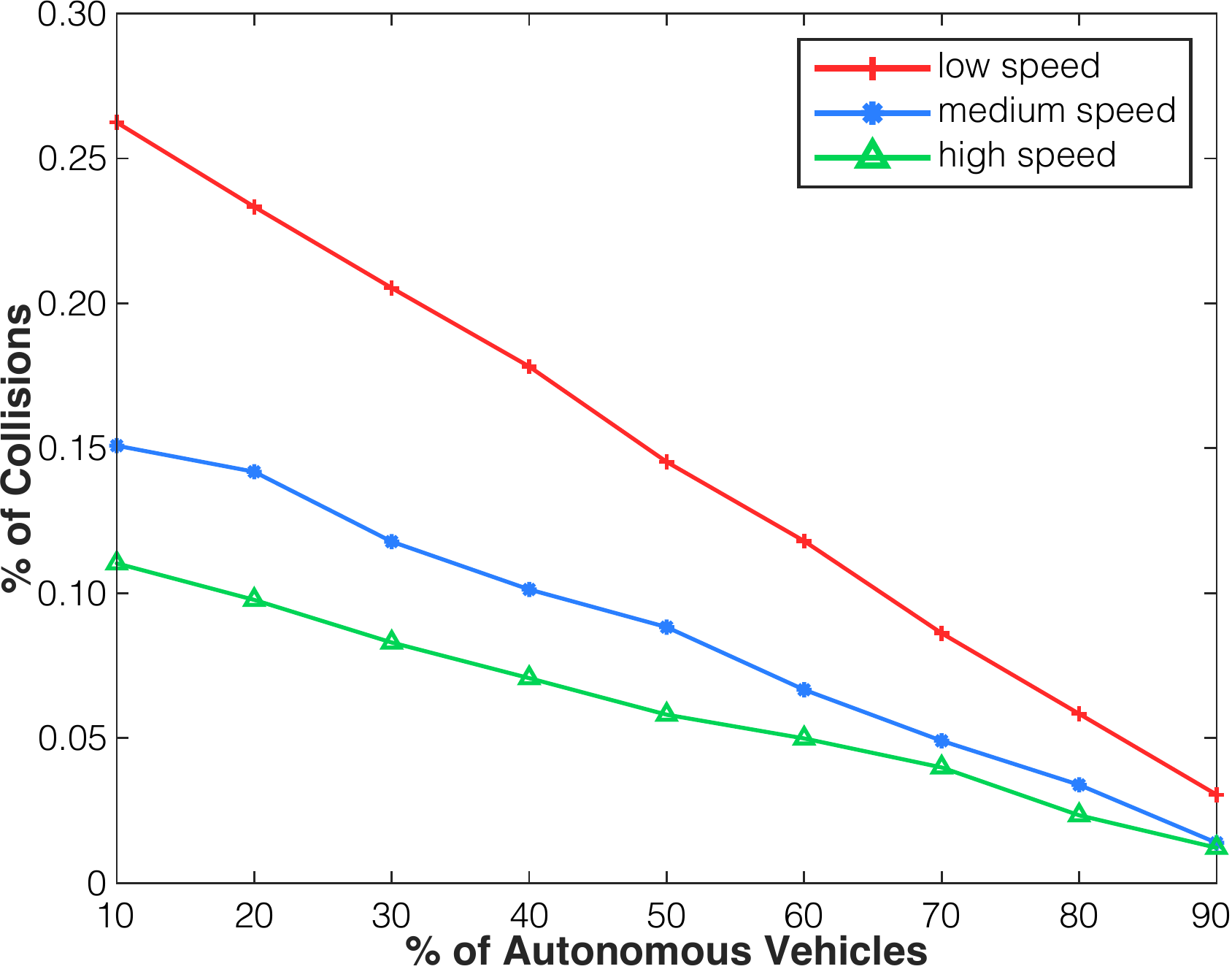}
\caption{Decreasing percentage of collisions of one crossing action at the increasing percentage of AVs.}
\label{fig:perc_collision}
\end{figure}

\begin{figure}
\centering
\includegraphics[width=\linewidth]{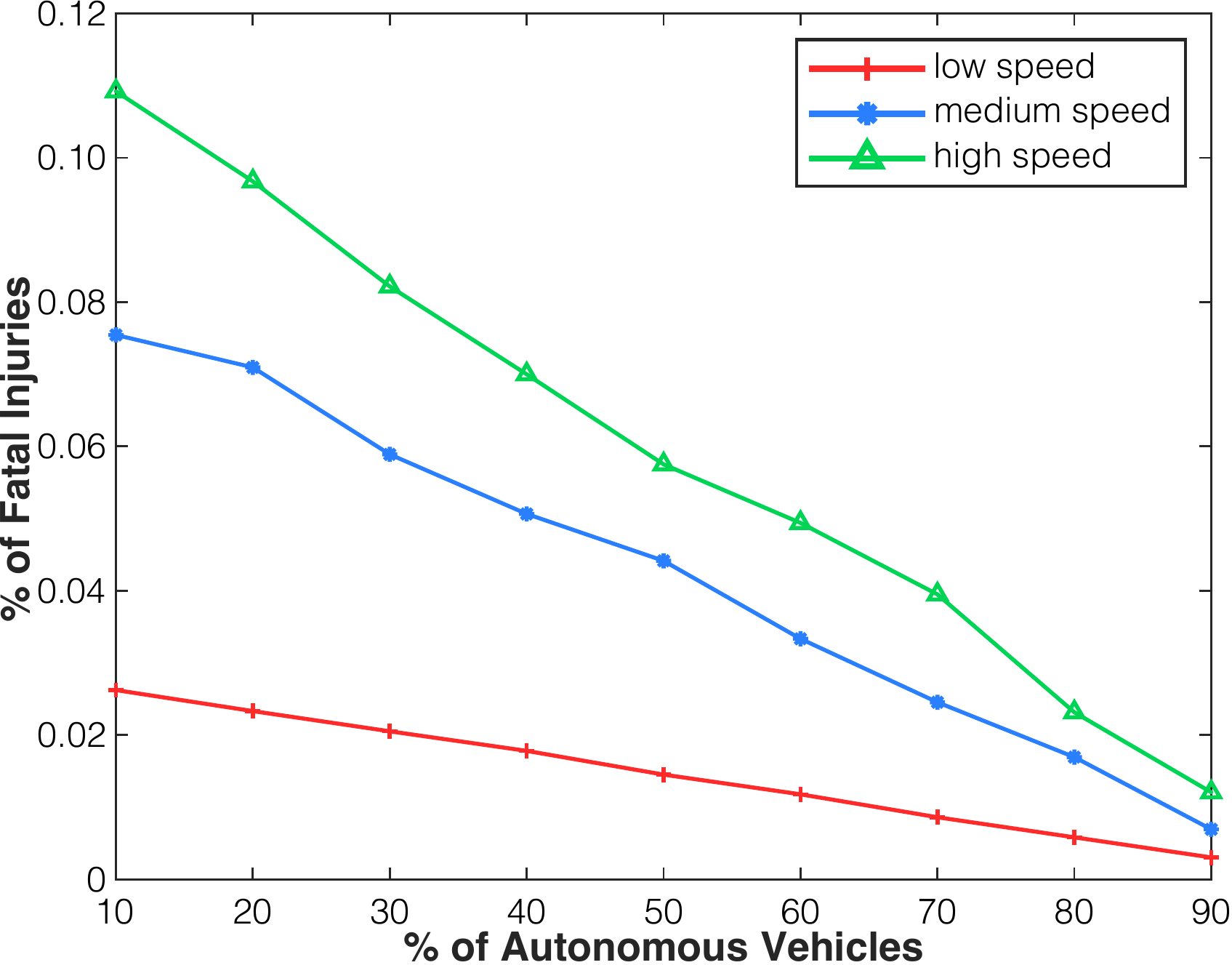}
\caption{Percentage of fatal injuries of one crossing action at the increasing percentage of AVs.}
\label{fig:perc_deaths}
\end{figure}

The results of a million iterations of the game are shown in Fig.\ \ref{fig:perc_collision}. Remarkably, the collision frequency exhibits a decreasing trend; while the exact slope of the curve depends on the evaluation of the outcomes; still, it may be worth investigating further in relation to different parameters.
Also, we can see that it is more likely to have incidents at lower speeds because pedestrians tend to cross the road more often due to the lower risk implicated. However, the death rate is much higher when higher speeds are involved. For example purposes we can think that the \textit{low} speed corresponds to about $30 \mathrm{\ km/h}$, the \textit{medium} speed to about $45 \mathrm{\ km/h}$ and the \textit{high} one to about $70 \mathrm{\ km/h}$. According to the report by the World Health Organization (WHO) in \cite{injuries} the rates of fatal injuries are approximately $10\%$ at $30 \mathrm{\ km/h}$, $50\%$ at $45 \mathrm{\ km/h}$ and more than $99\%$ at $70 \mathrm{\ km/h}$. With those assumptions the fatal injuries rate is reported in Fig.\ \ref{fig:perc_deaths}. These results obtained from game theory underline that a vehicle driving at high speed causes a lower number of accidents but a greater share causing fatal injuries.

The proposed model shows also that the AVs will always stop because they will act as safe as possible, thus the pedestrians and the cyclists will act more boldly: there will be no risk for them to cross in front of an AV, while with a human-driven vehicle the risk is much greater because the driver may be absent-minded or may not see the pedestrian crossing. 
The model also assumes that the cyclist can recognize an AV from a human-driven vehicle, which is sensible since AVs will likely be esthetically quite different from current vehicles. Thanks to this, the cyclist can determine whether to cycle or to mix between cycle and yield. With a low number of human-drivers in the traffic, then for the cyclist is prevailing to cycle.

If AVs end up to represent the majority of vehicles, it could be very likely that pedestrian and cyclists tend to dominate the scenario (i.e., they always win the contention for crossing the street).
To mitigate this behavior, some expedients can be used, for example crosswalks can be re-designed by city-planners also with physical barriers in order to reduce the direct interactions between pedestrians and AVs and, at the same time, street regulation could be changed so that pedestrians do not have priority when crossing the road, as pointed out in \cite{millard}. For example purposes, we can think of a fine for both cyclists and pedestrians crossing the roads without looking or waiting their right to cross. This would be possibly reflected into some payoffs variations, and a new game analysis.\\ 
For cyclists, the worst cases are when they cycle ($-600$) or walk ($-500$), respectively, and in both cases the AV drives; the values are lower than the previous ones because in this new setup the cyclist is also partially responsible for the accent. Both walk/stop and cycle/stop have been penalized with a fine of $15$, which is assumed to be a certain punishment and it is common knowledge: then yield/stop ($15$) and yield/go ($18$) are the right solutions according to the new street regulations.\\
For the AV, the worst outcomes are to hit the pedestrian ($-400$) or the cyclist ($-200$). If they both yield ($4$), they need to renegotiate the solution. Then walk/stop and cycle/stop ($5$) are the next worst outcomes because the AV had the right to pass. The best outcome is achieved when they play yield/go ($10$).\\
The resulting game in normal form is represented by Table \ref{tab:1b}. The unique NE of the game is yield/go since it is immediate to see that no player has anything to gain by changing only their own strategy, therefore in a fully autonomous environment the cyclists will yield and the AVs will go, thus validating the effectiveness of the changes.

\begin{table}[]
\centering
\large
\begin{tabular}{l|c|c|}
\cline{2-3}
                            & Go        & Stop \\ \hline
\multicolumn{1}{|l|}{Yield} & 18,10     & 15,4 \\ \hline
\multicolumn{1}{|l|}{Walk}  & -500,-400 & 0,5  \\ \hline
\multicolumn{1}{|l|}{Cycle} & -600,-200 & 5,5  \\ \hline
\end{tabular}
\caption{Normal form of the game in Section \ref{sec:res1} with modified payoffs.}
\label{tab:1b}
\vspace{-0.6cm}
\end{table}

\section{Pedestrian versus Vehicle using Entry Game} \label{sec:res2}
\begin{figure}
\centering
\includegraphics[width=\linewidth]{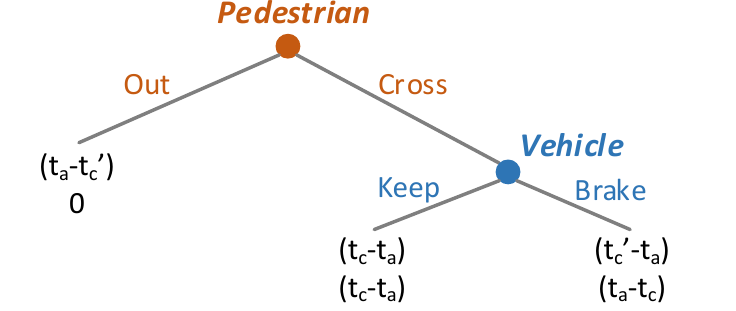}
\caption{Extensive form of the Entry Game of Section \ref{sec:res2}.}
\label{fig:222}
\end{figure}

A different approach is discussed in the following model: instead of focusing on the generic environment, in this section the attention is moved to the specific interaction between a pedestrian on a sidewalk, who have to decide whether to cross the road by jaywalking or not, and an upcoming vehicle.\\
The main differences from the previous model are: 
\begin{itemize}
\item the absence of zebra crossing and the unknown types,
\item actions are sequential,
\item payoffs are not axiomatically given, but computed as functions of environmental variables,
\item the game itself does not consider at the same time an AV and a human-driver. The comparison is made afterwards between the simulation results acquired from two instances of the same game with different parameters. 
\end{itemize}
To look from a different standpoint, we consider an \textit{Entry Game}, where the first move is made by the pedestrian, who decides whether to cross the road or to stay out on the sidewalk, and after that, the car decides between keep going or brake.
A possible payoff to consider is time, because a pedestrian approaching a crossing will consider, in first approximation, the estimated car arrival time to decide whether to cross or not. Fig.\ \ref{fig:222} shows the extensive form of the game.

In the proposed model, the pedestrian spends a constant time to cross the road, computed using the parameters indicated in \cite{speedwalk}, for what concerns the preferred speed-walk of $1.4 \mathrm{\ m/s}$, and in \cite{roadwidth}, for what regards the worst case lane width of $3.75 \mathrm{\ m}$ in many countries including Italy. The crossing time for the pedestrian is $t_a \approx 2.67 \mathrm{\ s}$. Moreover, the time taken by an AV in order to reach the trajectories-intersection point is
$$\begin{aligned}
& t_c' = \frac{\sqrt{v^2+2ad}-v}{a} \mathrm{\ \ \ in\ case\ of\ deceleration}&&\\\nonumber 
& t_c = \frac{d}{v}  \mathrm{\ \ otherwise}
\end{aligned}$$
where  $a$ is the acceleration, $d$ is the distance between vehicle and pedestrian, and $v$ is the car speed at distance $d$.


The payoff for the pedestrian, in case of crossing, is calculated as the difference between the time taken by the car to reach him ($t_c'$ or $t_c$ according to the cases) and his crossing time. It will be negative in case of collision and will increase as much as the pedestrian's perception of safety. On the other side, if the pedestrian decides to stay out gets a gain equal to the difference between his crossing time and the car approaching time (braking case, $t_c'$). This interval turns negative in case $t_c'$ is big enough to allow him crossing.\\
The payoff for the car, in case of a static pedestrian, is set a priori equal to 0, according to the absence of actions carried out by the vehicle. In the crossing scenario, instead, $t_c-t_a$ acts as a pivot: if it is greater than $0$ the car has no incentive to brake because it reaches later the intersection point, otherwise if it is lower than $0$ the vehicle has to brake, even if it is not sure to stop in time (this computation is demanded to the pedestrian before making a decision).
As a sequential game, it can be solved via backward induction (i.e. proceeding from the bottom of the tree and choosing what to do in any situation); in this way, the equilibrium discovered is not only a NE but also a Subgame Perfect Equilibrium (SPE), in fact with these payoffs settings the strategies chosen by the players always lead to a NE in each subgame. Since backward induction focuses only on pure strategies, the equilibrium is not influenced by the difference among them but only by their ordinality. Solving the game bring to three cases, that become NE according to the variables $d$ and $v$, as shown in Table \ref{tab:EntryGameNE}.

\begin{table}[]
\centering
\begin{tabular}{l|l|l|l|}
\cline{2-4}
                                & Case 1    & Case 2         & Case 3     \\ \hline
\multicolumn{1}{|l|}{Condition} & $t_a<t_c$ & $t_c<t_a<t_c'$ & $t_a>t_c'$ \\ \hline
\multicolumn{1}{|l|}{NE}        & CK        & CB             & O          \\ \hline
\end{tabular}
\caption{NE shift based on environmental variables. CK=Cross/Keep, CB=Cross/Brake, O=Out.\vspace{-0.8cm}}
\label{tab:EntryGameNE}
\end{table}

For the simulation, it is assumed that the AV, in case of braking, will start at constant deceleration $a = -2.5\ \mathrm{m/s}^2$ from the initial distance generated as a uniform random variable $d \sim \mathcal{U}(10,50) \mathrm{\ m}$; the speed value, instead, is assumed to be Gaussian distributed around at $30$ km/h,
i.e., $v \sim \mathcal{G}(30,10) \mathrm{\ km/h}$. The number of iterations has been set to $1000000$ and the aim is to show the distribution of the NE over the three different cases.

As a second experiment, the same interaction seen before but in case of a human driver is analyzed. In this situation, the braking action has to be revised due to the driver reaction time, the worst case analysis is considered setting it to $t_r = 1.5 \mathrm{\ s}$ \cite{brake} (accounting for both the processing time of the brain and the car dynamics but many other conditions may influence the result, depending on the physical model), thus the total approaching time for the car in case of deceleration becomes: 
$$ t_c'' = \frac{\sqrt{v^2+2a(d-d_r)}-v}{a} +t_r \mathrm{\ \ \ where} \ d_r= v \cdot t_r $$
Another modification needed in order to adapt the model to human behavior is the larger average speed, as human drivers tend not to always respect the road limits. The random variable representing speed is modified to $v \sim \mathcal{G}(50,10) \mathrm{\ km/h}$ and the same simulation is performed.

\begin{figure}
\centering
\vspace{0.3cm}
\includegraphics[width=\linewidth]{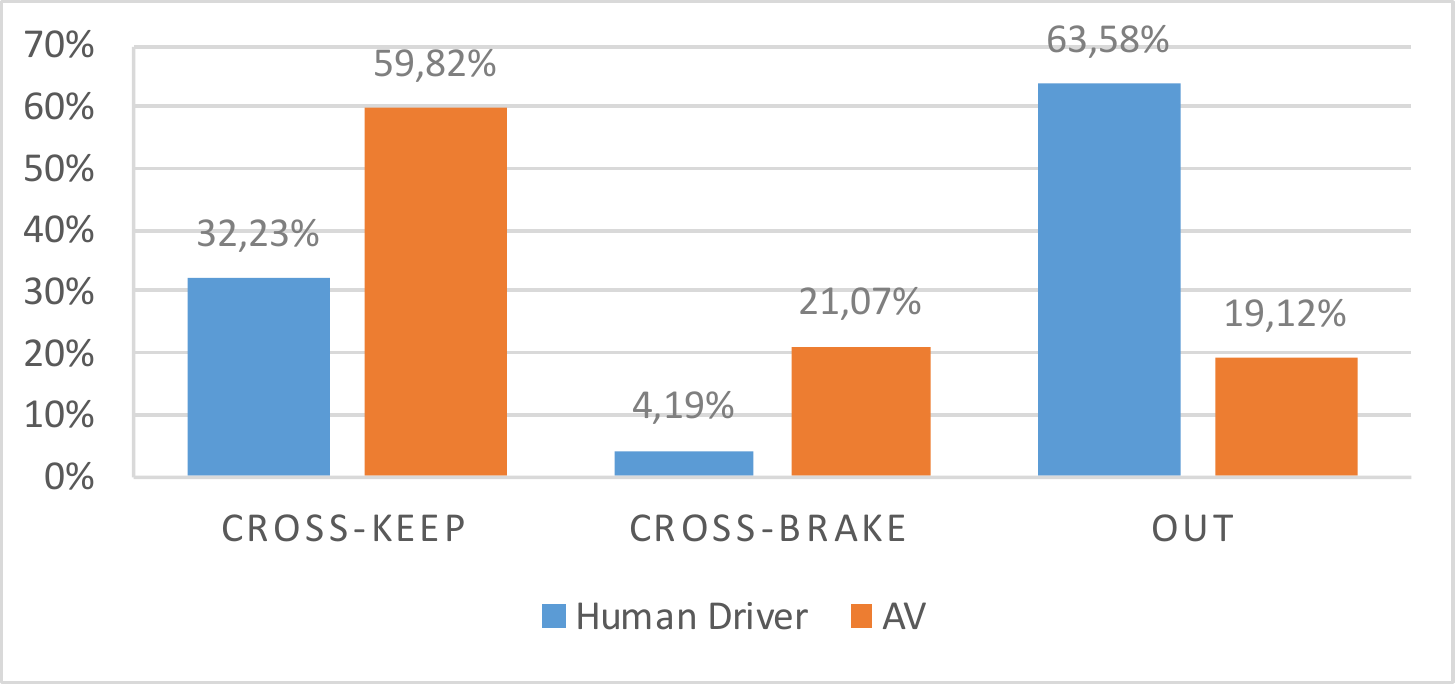}
\vspace{-0.8cm}
\caption{Histogram of occurrences for comparison between human-driven vehicle and autonomous vehicle.}
\label{fig:hist2}
\end{figure}

The results, summarized in Fig.\ \ref{fig:hist2}, show a decreasing number of the overall \textit{Cross} situations in case of human driver, while in case of AV there is an increasing number of such situations. In the case of AVs, the pedestrian will be not only safer but will tend to cross more frequently due to general lower speed of AVs and faster reaction time of machine-type drivers with respect to human-drivers. Moreover, since the AVs tend to slow down in presence of pedestrians, there is a higher number of situations of Cross-Keep.

However, the pedestrian may not properly know whether the human driver is alert or not and may consider a mean reaction time of about one second; in this setting if the driver is absent-minded may not be able to stop in time and the evaluation of the pedestrian may be wrong. This situation can be modeled considering an exponential reaction plus movement time starting from $0.8 \mathrm{\ s}$:  $t_r'\sim 0.8+\mathrm{Exp}(0.2) \mathrm{\ s}$. With these considerations, about  $3\%$ of the times the pedestrian implicitly assumes a lower reaction time than the actual one and $0.036\%$ of the times it ends in a road accident. \\
Notice however, as already stated, that this paper proposes some game theoretical models which can be useful for further analysis; all the values need to be set and supported by real data acquisitions but the baseline framework is truly reliable and reusable although very lightweight as regards the computational demand.

\section{Conclusions and Future Work}\label{sec:conclusion}
In this paper, some applications of game theoretical techniques have been proposed and discussed. The literature already reports that game theory is a useful application, and key tools are supplied to deal with strategic interactions, highlighting causal relationships among the road users. Nevertheless, models are often simplistic, since making the model too complex leads to a loss of physical meaning where the right strategy to consider is unclear. 

We applied a mixture of game theoretic techniques with some realistic and sensible assumptions to reach the following conclusions. The increasing number of AVs will reduce the number of collisions but also it will imply a new approach to road safety for pedestrians, arising the requirement of new regulations; moreover, communication systems among the AVs will be needed in order to optimize their traffic behavior. Finally, game theoretic analysis also has the advantage of allowing for distributed lightweight implementation inside the vehicles.

Our future interests regard a possible development of the scenarios presented in order to exploit also statistical tools, which can bring into our framework the real data that can validate the fairness and the calibration of the models.\\
As a follow-up study, for example, we intend to improve the accuracy of the model, by the means of deep investigation on the statistical road-user behaviors to set the payoffs more faithfully, as well as to perform adaptation on real-world data whenever they are available. Additionally, the simplistic assumption that the AVs always yield will be relaxed under specific conditions.

\bibliography{biblio}

\end{document}